\documentclass[journal]{IEEEtran}
\ifCLASSINFOpdf
\else
\fi
\usepackage{graphicx}      

\usepackage{amsmath} 
\usepackage{amssymb}  
\usepackage{graphicx}  
\usepackage{amsmath}

\usepackage{etoolbox}
\usepackage{float}
\usepackage[normalem]{ulem}
\useunder{\uline}{\ul}{}
\usepackage{url}
\usepackage{multicol}
\usepackage{subfigure}
\usepackage[export]{adjustbox}
\usepackage{cite}
\usepackage{graphicx}      

\begin{document}
%
\title{Optimal Scheduling of Multi-Energy Systems with Flexible Electrical and Thermal Loads}
%
%
%

\author{Ashok Krishnan,
	Bhagyesh V. Patil
        and Foo Y. S. Eddy,~\IEEEmembership{Member,~IEEE}
\thanks{Ashok Krishnan and Foo Y. S. Eddy are with the School of Electrical and Electronic Engineering, Nanyang Technological University, Singapore 639798. e-mail: (ashok004@e.ntu.edu.sg, eddyfoo@ntu.edu.sg).}
\thanks{Ashok Krishnan and Bhagyesh V. Patil are with the Cambridge Centre for Advanced Research and Education in Singapore Ltd., 1 Create Way, CREATE Campus, Singapore 138602. e-mail: (bhagyesh.patil@gmail.com).}}
\maketitle

\begin{abstract}
This paper proposes a detailed optimal scheduling model of an exemplar multi-energy system comprising combined cycle power plants (CCPPs), battery energy storage systems, renewable energy sources, boilers, thermal energy storage systems, electric loads and thermal loads.
The proposed model considers the detailed start-up and shutdown power trajectories of the gas turbines, steam turbines and boilers. Furthermore, a practical, multi-energy load management scheme is proposed within the framework of the optimal scheduling problem. The proposed load management scheme utilizes the flexibility offered by system components such as flexible electrical pump loads, electrical interruptible loads and a flexible thermal load to reduce the overall energy cost of the system. The efficacy of the proposed model in reducing the energy cost of the system is demonstrated in the context of a day-ahead scheduling problem using four illustrative scenarios.
\end{abstract}


\IEEEpeerreviewmaketitle

\section{Introduction}
%
%
%
%
\IEEEPARstart{N}{atural} gas plays an important role in the energy mix of many countries. For example, in 2015, 75\% of all the licensed electricity in Singapore was generated using CCPPs and tri-gen plants \cite{EMA15}. However, there is a growing clamour worldwide for increasing the percentage of renewable energy sources (RESs) in the energy mix. To mitigate the impact of intermittent RESs, numerous solutions have been discussed in the literature. Some of these solutions include the deployment of battery energy storage systems (BESSs) and open cycle gas turbines (GTs) apart from increasing the transmission capacity (see for instance \cite{7539382}, \cite{6006567}).
As such, future energy systems are widely expected to be heterogeneous in nature with the deployment of a wide array of technologies to deal with various operational scenarios.

Industrial parks such as Jurong Island in Singapore and the Yeosu National Industrial Complex in South Korea host numerous energy intensive industries \cite{KIM2010690}. In addition to this, many industrial parks including Jurong Island source most of their energy requirements from combined cycle power plants (CCPPs) \cite{EMA15}. CCPPs are power generating units which lower greenhouse gas emissions by producing both electricity and useful heat from a single fuel, usually natural gas. Typical CCPPs are known to exhibit efficiencies of up to 75\% \cite{Boyce01}. CCPPs are capable of operating in several modes, thereby offering a lot of flexibility to energy system operators. As such, many industrial complexes have the potential to be refurbished as eco-industrial parks (EIPs) wherein the CCPPs can be used to bridge the thermal and electrical energy streams \cite{KASTNER2015599},\cite{adhs}. In this scenario, the optimal management of multi-energy systems holds great promise in improving the overall energy efficiency of EIPs. 

In the context of multi-energy systems, there has been a lot of research interest in developing optimal scheduling models for CCPPs. Reference \cite{6039463} summarizes the various modelling approaches which have been used to model CCPPs by independent system operators (ISOs) such as ERCOT, PJM, NYISO and others. Among the approaches listed in \cite{6039463}, the physical unit modelling (component-based modelling) approach has been used by several researchers \cite{Kim05,4813189,adhs}. Some advantages of the component-based approach include the consideration of minimum on/off time, ramp limits, cost benefits and auxiliary equipment like boilers and duct burners in the scheduling model \cite{4813189}. 

The significant coupling which exists between the thermal and electrical energy streams makes the management of multi-energy systems a non-trivial problem \cite{6969121}. For instance, in a CCPP, the performance of the bottoming cycle always depends on the performance of the topping cycle. In recent years, there has been growing research interest in devising optimal management and operation strategies for multi-energy systems. A recent work considered the participation of a portfolio of generators comprising wind farms and combined heat and power (CHP) plants in the Nordic two-price balancing market \cite{7124532}. Specifically, many researchers have focused on developing optimal scheduling problem formulations for multi-energy systems in general and microgrids in particular. The authors of \cite{6969121} formulated and solved an optimal scheduling problem for combined cooling, heating and power (CCHP) plants to satisfy electrical and cooling loads in a microgrid scenario. In \cite{LI2018974}, the authors examined the coordinated scheduling of microturbines and other distributed generators to satisfy electrical, thermal and cooling loads in grid-connected and islanded microgrids. A multi-energy demand response program for the optimal management of energy hubs including CHP plants was proposed in \cite{7987711}. The authors of \cite{8372457} proposed a robust optimization framework for handling power market price uncertainties in CHP-based multi-energy microgrids. An earlier work proposed an optimization model for a grid-connected microgrid comprising several residential micro CHPs \cite{KOPANOS20131522}. The framework proposed in \cite{KOPANOS20131522} also permitted the interchange of electrical and thermal power between subgroups of generators within the overall microgrid. A recent work proposed a multistage stochastic mixed integer linear programming (MILP) framework for the participation of a CHP plant with heat storage in multiple, sequential electricity markets wherein the stochastic processes were used to capture price uncertainties \cite{KUMBARTZKY2017390}. 


The optimal scheduling of larger multi-energy systems such as those found in industrial parks and shipyards has been less explored by researchers. A few examples of such formulations can be found in \cite{Kim05,4813189,adhs,KIA2017241}. A mixed integer nonlinear programming (MINLP) formulation for the multi-energy scheduling problem in a university campus was presented in \cite{Kim05}. Reference \cite{Kim05} considered a detailed, component-wise scheduling model of the CHP plant including the start-up (SU) and shutdown (SD) power trajectories of the gas turbines (GTs), steam turbines (STs) and boilers. The authors of \cite{4813189} proposed an approximated mixed integer programming (MIP) formulation of the multi-energy scheduling problem. The system considered in \cite{4813189} comprised both CCPPs and conventional thermal units. The authors' recent work in \cite{adhs} combined elements from their previous works in \cite{Asho01,KRISHNAN20179329} apart from \cite{4813189,Kim05} to develop a mixed integer quadratic programming (MIQP) formulation for the multi-energy scheduling problem. Furthermore, \cite{adhs} demonstrated the potential of pump scheduling optimization (PSO), an industrial load management technique, in reducing the overall energy cost for the system operator. In \cite{KIA2017241}, the authors proposed an optimal, day-ahead scheduling problem formulation including security constraints for CHP-based multi-energy systems comprising CHPs, boilers, battery energy storage systems (BESSs) and thermal energy storage systems (TESSs).

Conventional optimal power system scheduling problem formulations ignore the startup/shutdown power trajectories which are intrinsic to large generators. Consequently, the optimal scheduling problem does not allocate a large amount of energy which is actually present in real-time, thereby distorting the actual load balance and system reserve requirements \cite{Morales-España2015,MORALESESPANA2017223}. Ignoring the startup/shutdown power trajectories could thus lead to inefficiencies in the operation of the power system and economic losses \cite{6365287}. While the pitfalls of ignoring the startup/shutdown power trajectories are well known, they continue to be ignored in many scheduling problem formulations owing to the complexities involved in solving the resulting optimization problem. However, the authors note that many multi-energy systems such as EIPs are usually smaller than conventional power systems in terms of the number of generators. Furthermore, the startup/shutdown trajectories are also an intrinsic part of boilers which are important components of multi-energy systems. Computationally efficient approaches for handling the startup/shutdown power trajectories in optimal scheduling problems have also been proposed recently \cite{6365287}. 

The authors note that \cite{6969121,7124532,LI2018974,7987711,8372457,4813189,KIA2017241,KOPANOS20131522,KUMBARTZKY2017390} do not model the startup/shutdown power trajectories for the CCPP/CHP plants and boilers. While \cite{Kim05} and the authors' recent work in \cite{adhs} considered the startup/shutdown power trajectories for the CCPPs and boilers, they did not examine the interactions between the CCPPs/boilers and the other multi-energy system components such as BESS, TESS, renewable energy sources (RESs) and flexible electrical and thermal loads. Furthermore, multi-energy load management strategies were included within the framework of an optimal multi-energy scheduling problem recently in \cite{7987711}. However, the multi-energy load management strategy presented in \cite{7987711} was generic in nature without considering any specific industrial load management application. Consequent to the above discussions, this paper proposes a comprehensive optimal scheduling problem formulation for an exemplar multi-energy system comprising CCPPs, boilers, RESs, BESS, TESS, flexible electrical pump loads, electrical interruptible loads (ILs) and a flexible thermal load. Based on the above discussions and compared to the existing works in the literature, the major contributions of this paper are summarized below: 
\begin{enumerate}
	\item A detailed optimal scheduling model of an exemplar multi-energy system is developed which considers the startup/shutdown power trajectories which are inherent to the CCPPs (GTs and STs) and boilers. Three startup methods (hot, warm and cold) are modelled for each GT, ST and boiler. Each startup method has a unique cost associated with it. This work also examines the optimal coordinated operation of the CCPPs, boilers, RESs, BESS, TESS, ILs and flexible electrical and thermal loads to meet the thermal and electrical load demands in the system.	   
	\item A multi-energy load management scheme is proposed including a practical industrial pump scheduling problem. Furthermore, the proposed load management scheme also utilizes the flexibility offered by system components such as the ILs, the flexible pump loads and the flexible thermal load.
\end{enumerate}
The efficacy of the proposed optimal scheduling problem formulation is demonstrated using illustrative numerical case studies.

The remainder of this paper is organized as follows: Section II describes the development of the scheduling model of each component in the multi-energy system considered in this work. Furthermore, the integration of the individual component models to form the system model using the MLD framework is also described in Section II. The optimal scheduling problem for the multi-energy system is formulated in Section III. Section IV presents the results of the numerical case studies performed to demonstrate the efficacy and the utility of the optimization model developed in this paper. Finally, some concluding remarks are presented in Section V. 

\section{System Model}
\begin{figure}[h!]
	\begin{center} \includegraphics[height=2.5in]{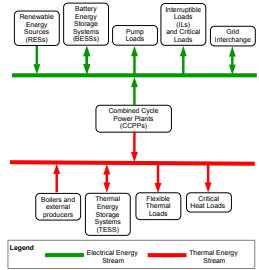}
		\caption{Overview of an exemplar multi-energy system}  \label{overview}
	\end{center}
\end{figure}

This section describes the various components of the multi-energy system considered in this paper. An overview of the exemplar multi-energy system considered in this paper is shown in Fig. \ref{overview}. As shown in Fig. \ref{overview}, the CCPPs act as bridges between the electrical and thermal energy streams in the multi-energy system. The RESs produce only electrical energy while the boilers produce only thermal energy. The BESS and TESS can produce and consume electrical and thermal energy respectively. Apart from this, as shown in Fig. \ref{overview}, the multi-energy system also contains different types of loads which only consume energy.
The multi-energy system considered in this paper comprises 2 CCPPs (each comprising 1 GT and 1 ST), 2 boilers, a BESS, 2 wind power plants (RESs), 2 TESSs, flexible industrial pump loads, a flexible thermal load and ILs. The electrical power system is also enabled to exchange (buy/sell) power with the main utility grid. There is also an option to purchase thermal energy from external producers to fulfil the thermal load demand.

\subsection{CCPP Components}

\begin{figure}[h!]
	\begin{center} \includegraphics[width=3.3in, height=2in]{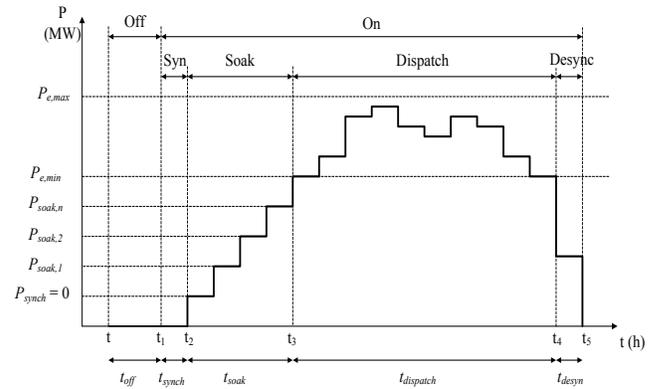}
		\caption{Typical start-up and shutdown power trajectories of a thermal unit}  \label{fig1}
	\end{center}
\end{figure}

Each CCPP considered in this paper comprises 1 GT, 1 ST and 1 heat recovery steam generator (HRSG). Additionally, 1 boiler and 1 TESS are associated with each CCPP. Due to the presence of two thermodynamic cycles (Brayton cycle for the GT and Rankine cycle for the ST), the overall energy efficiency of a CCPP is 20-30\% higher than a traditional, single cycle thermal power plant. 
In a CCPP, the HRSG is used to recover the waste heat emitted by the GT. The HRSG acts as a heat exchanger between the two thermodynamic cycles. The output of the HRSG is high pressure steam. To augment the HRSG steam output during periods of high thermal load demand, a boiler is used to generate steam.
The different operating modes of the GTs, STs and boilers are illustrated in Fig. \ref{fig1}. In electrical power systems, the ramping constraints are broadly classified into three types: 1) Operating ramp constraint, 2) Start-up ramp constraint and 3) Shutdown ramp constraint \cite{Arr02}. 
The start-up ramp constraint refers to a predefined trajectory during the unit start-up wherein the electrical power output from the unit gradually increases to its technical minimum level. In this work, three start-up methods (hot, warm and cold) are modelled for each GT, ST and boiler. 
Each GT, ST and boiler model is designed to select the correct start-up method depending on the prior downtime \cite{Asho01}. A unique electrical output power trajectory is specified for each start-up method in the GT and ST models. 
The shutdown ramp constraint refers to a predefined trajectory during the unit shutdown wherein the electrical power output first reduces to the technical minimum level before reducing to 0MW. 
Furthermore, it is assumed that the boilers do not produce any thermal power during the start-up and shutdown processes.

As shown in Fig. \ref{fig1}, each unit may typically operate in four distinct phases - synchronization phase, soak phase, dispatch phase and desynchronization phase (\cite{Kim05}, \cite{Mor03}, \cite{Sim04}). The start-up trajectory is associated with the synchronization and soak phases while the shutdown trajectory is associated with the desynchronization phase.

For each GT, ST and boiler, the exemplar start-up trajectory shown in Fig. \ref{fig1} illustrates that the time required to enter the dispatch phase increases as the downtime prior to commitment increases.
This is essential to avoid any mechanical stresses.
After synchronization with the grid (synchronization phase), STs enter the soak phase. GTs enter the soak phase on being committed. The electrical power output of a GT or ST during the soak phase may increase linearly in steps to its technical minimum levels. Fig. \ref{fig1} illustrates a generalized scenario wherein $P_{\textrm{soak,1}}$, $P_{\textrm{soak,2}}$...$P_{\textrm{soak,n}}$ represent the different electrical power outputs of a unit during the different stages of the soak phase respectively. In this work, it is assumed that a constant electrical power, $P_{\textrm{soak},k}^{f}$ is produced by a unit $f$ in the soak phase during hour $k$. The soak phase is followed by the dispatch phase wherein the unit operates between its technical minimum and maximum electrical power outputs. Similarly, during shutdown, a unit first undergoes the desynchronization phase. Subsequently, the electrical power output of the unit drops to zero.

Each boiler also passes through the soak phase while being started up. 
The duration of the soak phase determines the time required by a boiler to reach the dispatch phase on being committed. As detailed later in this section, the scheduling model of the boiler considers the soak phase duration. Since the boilers do not need to synchronize and desynchronize from the utility grid, they do not undergo the synchronization and desynchronization phases. 
The mathematical scheduling models of the GTs, STs and boilers are presented in the following paragraphs. 

\subsubsection{Minimum up/down time constraint}
The binary input variable $u^f_{k}$ is set to 1 if unit $f$ enters the dispatch phase during the time interval [$k-UT,k$]. Conversely, $u^f_{k}$ is set to 0 if unit $f$ enters the desynchronization phase during the interval [$k-DT,k$]. $UT$ and $DT$ represent the minimum uptime and minimum downtime parameters respectively. The parameters $UT$ and $DT$ are both set at 3 hours for all the GTs, STs and boilers considered in this paper. For a unit $f$, the minimum uptime and minimum downtime constraints are formulated as shown in \eqref{uptime} and \eqref{downtime} respectively:
\begin{align}
\label{uptime}
\sum_{\tau=k}^{k+UT-1}u^f_{\tau} \geq UT[w_{\textrm{disp},k}^{f} - w_{\textrm{disp},k-1}^{f}],~\forall k\in \mathcal{K},\nonumber \\ \forall f\in \{GT,ST,BR\}\\
\label{downtime}
\sum_{\tau=k}^{k+DT-1}[1 - u^f_{\tau}] \geq DT[w_{\textrm{shutdown},k-1}^{f} - w_{\textrm{shutdown},k}^{f}],\nonumber\\\forall k\in \mathcal{K}, \forall f\in \{GT,ST,BR\}    
\end{align}
where $u^f_{k}$ represents the commitment status of unit $f$ during hour $k$; $GT=\{\textrm{GT1, GT2}\}$, $ST=\{\textrm{ST1, ST2}\}$ and $BR=\{\textrm{Boiler 1, Boiler 2}\}$ represent the sets of GTs, STs and boilers in the system respectively; $\mathcal{K}$ represents the set of all the hours in a day i.e. $\mathcal{K}=\{1,2,\ldots,24\}$; $w_{\textrm{disp},k}^{f}$ is a binary auxiliary variable which is set to 1 if the dispatch phase of unit $f$ commences during hour $k$ and $w_{\textrm{shutdown},k}^{f}$ is a binary auxiliary variable which is set to 1 if the shutdown phase of unit $f$ commences during hour $k$. 
%
\subsubsection{Start-up type selection}
While scheduling each unit $f$, it is important to ensure that the correct start-up method is selected based on the prior downtime of the unit. The following equations are used to select the appropriate start-up method:
\begin{equation}
w_{\textrm{start-up},k}^{f} \leq \sum_{n \in \mathcal{N}}w_{\textrm{start-up},k}^{n,f}, \forall k\in \mathcal{K}, \forall f\in \{GT,ST,BR\}\\
\end{equation}
\begin{eqnarray}
\label{susel}
w_{\textrm{start-up},k}^{n,f} \leq \sum_{\tau = k-t_{\textrm{u}}^{n,f} + 1}^{k-t_{\textrm{l}}^{n,f}}w_{\textrm{shutdown},\tau}^{f},\nonumber \\~\forall k\in \mathcal{K}, \forall f\in \{GT,ST,BR\},\forall n \in \mathcal{N} 
\end{eqnarray}
where $\mathcal{N} = \{\textrm{cold, warm, hot}\}$ represents the set of start-up methods; $w_{\textrm{start-up},k}^{n,f}$ is a binary auxiliary variable which is set to 1 if start-up method $n$ of unit $f$ is initiated during hour $k$. Equation \eqref{susel} ensures that this is possible only if the shutdown process of unit $f$ was initiated during the time interval $[k-t_{\textrm{u}}^{n,f},k-t_{\textrm{l}}^{n,f}]$. Finally, $w_{\textrm{start-up},k}^{f}$ is a binary auxiliary variable which is set to 1 if unit $f$ is either in the synchronization phase or the soak phase of any start-up method during hour $k$.
%

%

\subsubsection{Synchronization and Soak Phases}
The synchronization phase of start-up method $n$ is identified as shown below:
\begin{equation}
w_{\textrm{synch},k}^{n,f} = \sum_{\tau = k-t_{\textrm{synch}}^{n,f} + 1}^{k}w_{\textrm{start-up},\tau}^{n,f},~\forall k\in \mathcal{K}, \forall f\in \{ST\}, \forall n \in \mathcal{N}
\end{equation}
where $w_{\textrm{synch},k}^{n,f}$ is a binary auxiliary variable which is set to 1 if ST $f$ is in the synchronization phase of start-up method $n$ during hour $k$ and $t_{\textrm{synch}}^{n,f}$ is the synchronization phase duration of start-up method $n$.

The soak phase of start-up method $n$ is identified as shown below:
\begin{eqnarray}
\label{soak}
w_{\textrm{soak},k}^{n,f} = \sum_{\tau = k - t_{\textrm{synch}}^{n,f} - t_{\textrm{soak}}^{n,f} + 1}^{k - t_{\textrm{synch}}^{n,f}}w_{\textrm{start-up},\tau}^{n,f},\nonumber \\~\forall k\in \mathcal{K}, \forall f\in \{GT,ST,BR\},\forall n \in \mathcal{N} 
\end{eqnarray}
where $w_{\textrm{soak},k}^{n,f}$ is a binary auxiliary variable which is set to 1 if unit $f$ is in the soak phase of start-up method $n$ during hour $k$ and $t_{\textrm{soak}}^{n,f}$ is the soak phase duration of start-up method $n$. 

\subsubsection{Desynchronization Phase Constraints}
The desynchronization phase of unit $f$ is identified as follows:
\begin{equation}
w_{\textrm{desyn},k}^f = \sum_{\tau = k + 1}^{k + t_{\textrm{desyn}}^f}w_{\textrm{off},\tau}^f~\forall k\in \mathcal{K},\forall f\in \{GT,ST\}  
\end{equation}
where $w_{\textrm{desyn},k}^f$ is a binary auxiliary variable which is set to 1 if unit $f$ is in the desynchronization phase during hour $k$; $w_{\textrm{off},k}^f$ is a binary auxiliary variable which is set to 1 if the electrical power produced by unit $f$ drops to 0MW during hour $k$ and $t_{\textrm{desyn}}^f$ is the desynchronization phase duration of unit $f$. 

\subsubsection{Reserve Constraints}
The spinning reserve constraints (electrical) for the system considered in this paper are defined as follows:
\begin{eqnarray}
(50 - P_{\textrm{eb},k})  + \sum_{\substack{f\in \{GT,ST\}}}SR^f_{k}x_{\textrm{disp},k}^f \geq SR_{k},~\forall k\in \mathcal{K} \\
SR^f_{k}x_{\textrm{disp},k}^f \leq 10MSR^{f},~\forall k\in \mathcal{K}, \forall f \in \{GT,ST\}\\
SR^f_{k}x_{\textrm{disp},k}^f + P^f_{\textrm{e},k}x_{\textrm{disp},k}^f \leq P^f_{\textrm{e,max}},\nonumber \\~\forall k\in \mathcal{K}, \forall f \in \{GT,ST\}
\end{eqnarray}
where $P_{\textrm{eb},k}$ is the electrical power purchased from the main utility grid during hour $k$ in MW whose upper bound is 50MW; $x_{\textrm{disp},k}^f$ is a binary state variable which is set to 1 if unit $f$ is in the dispatch phase during hour $k$; $SR^f_{k}$ is the spinning reserve contributed by unit $f$ during hour $k$; $SR_{k}$ is the total system spinning reserve requirement during hour $k$; $MSR^{f}$ is the maximum spinning rate of unit $f$ in MW/min; $P^f_{\textrm{e},k}$ is the electrical power (real power) produced by unit $f$ during hour $k$ in MW and $P^f_{\textrm{e,max}}$ is the upper bound on the electrical power produced by unit $f$ in MW.


\subsubsection{Ramping Constraints in Dispatch Phase}
Ramping constraints limit the electrical power outputs from the STs as shown below. The GTs are not subjected to this constraint due to their fast ramping capabilities. Furthermore, there are no ramping constraints on the production of thermal energy by the boilers in the dispatch phase.
\begin{eqnarray}
-0.5P_{\textrm{e,max}}^{f} \leq P^f_{\textrm{e},k}x_{\textrm{disp},k}^f - P^f_{\textrm{e},k-1}x_{\textrm{disp},k-1}^f \leq 0.5P_{\textrm{e,max}}^{f},\nonumber\\\forall k \in K, \forall f \in ST
\end{eqnarray}

\subsubsection{Thermal Power Generation Constraints}
In a CCPP, the performance of the topping cycle influences the performance of the bottoming cycle. The boiler associated with each CCPP is used to supplement the waste heat recovered by the HRSG. The total steam generated by each CCPP-boiler pair is either utilized by the corresponding ST to generate electricity or utilized to service thermal loads via a heat distribution network. Any excess steam which is generated may either be stored in the corresponding TESS for future use or emitted to the surrounding environment.
\begin{eqnarray}
P_{\textrm{h},k}^f = a_{\textrm{0}}^fP_{\textrm{e},k}^fx_{\textrm{disp},k}^f + a_{\textrm{1}}^f,~\forall k \in K, \forall f \in GT\\ 
P_{\textrm{h},k}^f = b_{\textrm{0}}^fw_{\textrm{br},k}^f,~\forall k \in K,~\forall f \in BR\\
h_k^f = b_{\textrm{1}}^fP_{\textrm{e},k}^f + b_{\textrm{2}}^f,~\forall k \in K,~\forall f \in ST\\
P_{\textrm{h},k}^{\textrm{GT1}} + P_{\textrm{h},k}^{\textrm{Boiler 1}} \geq h_k^{\textrm{ST1}}\\
P_{\textrm{h},k}^{\textrm{GT2}} + P_{\textrm{h},k}^{\textrm{Boiler 2}} \geq h_k^{\textrm{ST2}}
\end{eqnarray} 
where $P_{\textrm{h},k}^f$ is the thermal power produced by unit $f$ during hour $k$ in MW; $w_{\textrm{br},k}^f$ is the fuel consumed by boiler $f$ during hour $k$ in mcf; $a_{\textrm{0}}^f$ and $a_{\textrm{1}}^f$ are the constant coefficients of the electrical power - thermal power curve for GT $f$; $b_{\textrm{0}}^f$ is a conversion factor which relates the fuel consumed by boiler $f$ to its thermal power production; 
$h_k^f$ is the steam consumed by ST $f$ during hour $k$ in MW. Finally, $b_{\textrm{1}}^f$ and $b_\textrm{2}^f$ are the constant coefficients of the electrical power - thermal power curve for ST $f$. The following parameter values are used in this work: $a_{\textrm{0}}^{\textrm{GT1}}$ = 1.35, $a_{\textrm{1}}^{\textrm{GT1}}$ = 97.09; $a_{\textrm{0}}^{\textrm{GT2}}$ = 1.14, $a_{\textrm{1}}^{\textrm{GT2}}$ = 96.32; $b_{\textrm{0}}^{\textrm{BR1}}$ = 0.0004; $b_{\textrm{0}}^{\textrm{BR2}}$ = 0.0003; $b_{\textrm{1}}^{\textrm{ST1}}$ = 1.74, $b_{\textrm{2}}^{\textrm{ST1}}$ = 72.05; $b_{\textrm{1}}^{\textrm{ST2}}$ = 0.82, $b_{\textrm{2}}^{\textrm{ST2}}$ = 85.58.

\subsection{Battery Energy Storage System}

A realistic BESS model including intertemporal state-of-charge (SOC) constraints and operational limits is considered in this paper. Additionally, the BESS model includes a battery degradation cost which reflects the BESS purchase cost based on its utilization (charging and discharging). The overall BESS model is described below \cite{Kalpesh}.
\begin{eqnarray}
\label{e1}
SOC_{k+1} = SOC_{k} + (\eta_{\textrm{c}}P_{\textrm{bc},k} - P_{\textrm{bd},k}/\eta_{\textrm{d}})/P_{\textrm{1C}},~\forall k \in K\\
\label{e2}
SOC_{\textrm{min}} \leq SOC_{k+1} \leq SOC_{\textrm{max}},~\forall k \in K\\
\label{e3}
0 \leq P_{\textrm{bc},k} \leq P_{\textrm{bc,max}},~\forall k \in K\\
\label{e4}
0 \leq P_{\textrm{bd},k} \leq P_{\textrm{bd,max}},~\forall k \in K
\end{eqnarray}
The cost incurred due to the operation of the BESS is calculated as follows:
\begin{equation}
\label{e5}
C_{\textrm{BESS}} = \sum_{\substack{k\in \mathcal{K}}} \frac{I}{2B_{\textrm{cap}}N}(\frac{P_{\textrm{bc},k}}{T_{\textrm{bc}}}+\frac{P_{\textrm{bd},k}}{T_{\textrm{bd}}})
\end{equation}
where $P_{\textrm{bc},k}$ and $P_{\textrm{bd},k}$ are the charging and discharging powers of the BESS during hour $k$ respectively; $P_{\textrm{1C}}$ is the power required by the BESS to charge $100\%$ in 1 hour i.e. $\textrm{1C}$ rate; $(.)_{\textrm{min}}$ and $(.)_{\textrm{max}}$ represent the minimum and maximum bounds of the corresponding parameter respectively; $N$ represents the lifetime of the BESS in hours; $T_{\textrm{bc}}$ and $T_{\textrm{bd}}$ are the average number of hours the BESS charges and discharges in a day respectively; $\eta_{\textrm{c}}$ and $\eta_{\textrm{d}}$ are the charging and discharging efficiencies of the BESS respectively; $I$ is the capital cost of purchasing the BESS in \$/kWh and $B_{\textrm{cap}}$ is the capacity of the BESS in \textrm{kWh}. The SOC of the BESS evolves according to (\ref{e1}). Equations (\ref{e2}) - (\ref{e4}) represent constraints on the evolution of the BESS SOC, charging power and discharging power respectively. The parameters of the BESS used in this paper are as follows: $N$ = 6,000h, $P_{\textrm{bc,max}}$ = 7,386.645kWh, $P_{\textrm{bd,max}}$ = 7,615.095kWh, $\eta_{\textrm{c}}$ = $\eta_{\textrm{d}}$ = 0.97, $P_{\textrm{1C}}$ = 3.73MW*15 = 55.965MW, $SOC_{\textrm{min}}$ = 0.2 and $SOC_{\textrm{max}}$ = 0.8.

To the best of the authors' knowledge, a BESS with 30MW power capacity is not available as a single commercial system for ready deployment. However, BESSs with 2MW power capacity and 3.7MWh energy capacity are available in the market \cite{LG}. With the help of series-parallel combinations of such BESSs, a multi-modular BESS with 30MW capacity can be realized. Similar systems can be found installed at several locations \cite{Dayton}. Based on recent quotations obtained for such grid scale BESSs, the cost of the BESS used in this paper is estimated to be \$450/kWh.

\subsection{Renewable Energy Sources}
The multi-energy system considered in this paper contains two wind power plants which produce only electrical energy. The electrical power output of a wind turbine is proportional to $v_{\textrm{wind}}^{\textrm{3}}$ wherein $v_{\textrm{wind}}$ represents the wind velocity. The electrical power output of a wind turbine is calculated using the following equation \cite{WANG2015324}:
\begin{equation}
P_{\textrm{wind}} = 0.5C_{\textrm{p}}k\rho A (v_{\textrm{wind}})^{\textrm{3}}
\end{equation}
where $C_{\textrm{p}}$ represents the power coefficient which is a function of the tip speed ratio; $\rho$ represents the air density and $A$ represents the area swept by the rotor blades. For this paper, the generation forecasts of the wind power plants were obtained from \cite{ninja}. 

\subsection{Thermal Energy Storage System}
Accumulator tanks are thermal energy storage systems (TESSs) with high levels of insulation. Their operation is analogous to that of the BESSs which are used to store electricity. The discrete time, state-space model of a TESS is expressed as follows:
\begin{equation}
\label{h1}
H_{k+1}^{p} = H_{k}^{p} + Q_{\textrm{in},k}^{p} - Q_{\textrm{out},k}^{p} - \gamma_{k}^{p},~\forall k\in \mathcal{K},~\forall p \in \mathcal{P}  
\end{equation}
where $H_{k}^{p}$ is a continuous state variable which represents the storage level of TESS $p$ during hour $k$; $\mathcal{P}$ represents the set of all the TESSs in the system; $Q_{\textrm{in},k}^{p}$ represents the thermal power supplied to TESS $p$ during hour $k$; $Q_{\textrm{out},k}^{p}$ represents the thermal power supplied by TESS $p$ during hour $k$ and $\gamma_{k}^{p}$ represents the psychological discharge of TESS $p$ during hour $k$.
The operation of each TESS $p$ is constrained by the following:
\begin{eqnarray}
\label{h2}
H_{\textrm{min}}^{p} \leq H_{k}^{p} \leq H_{\textrm{max}}^{p},~\forall k\in \mathcal{K},~\forall p \in \mathcal{P}  \\
\label{h3}
0 \leq \gamma_{k}^{p} \leq \gamma_{\textrm{max}}^{p},~\forall k\in \mathcal{K},~\forall p \in \mathcal{P}  \\
\label{h4}
Q_{\textrm{in},k}^{\textrm{1}} \leq P_{\textrm{h},k}^{\textrm{GT1}} + P_{\textrm{h},k}^{\textrm{Boiler 1}} - h_k^{\textrm{ST1}},~\forall k\in \mathcal{K} \\
\label{h6}
Q_{\textrm{in},k}^{\textrm{2}} \leq P_{\textrm{h},k}^{\textrm{GT2}} + P_{\textrm{h},k}^{\textrm{Boiler 2}} - h_k^{\textrm{ST2}},~\forall k\in \mathcal{K}
\end{eqnarray}
In this work, two identical TESSs are modelled with the following parameter values: $H_{\textrm{min}}^{p}$ = 90MW; $H_{\textrm{max}}^{p}$ = 200MW and $\gamma_{\textrm{max}}^{p}$ = 20MW.

\subsection{Flexible Pump Loads}
Some industrial electrical loads can be scheduled to operate in a manner which reduces the total electricity cost of the system. In this work, large pump loads are modelled as exemplar flexible industrial (electrical) loads. The flexible pump loads allow the system operator to take advantage of lower electricity prices during certain hours of the day. The flexible pump loads also aid in eliminating or reducing uncontracted capacity and its associated cost. The operation of the flexible pump loads is constrained by the following:
\begin{eqnarray}
\label{pump1}
\sum_{\substack{k\in \mathcal{K}\\m\in \mathcal{M}}}Q^mu_k^m\geq V_{\textrm{d}}
\end{eqnarray}
where $\mathcal{M}$ represents the set of flexible pump loads in the system; $Q_k^m$ represents the flow rate of pump $m$ during hour $k$; $u_k^m$ represents the commitment status of pump $m$ during hour $k$ and $V_{\textrm{d}}$ is the total volume of liquid to be pumped in 24 hours.

Furthermore, due to their large inertias, large pumps cannot be started up and shut down too frequently. The maximum number of start-up and shutdown events permitted during a 24-hour period for pump $m$ is constrained as follows: 
\begin{eqnarray}
\label{pump2}
\sum_{k\in \mathcal{K}}w_{\textrm{SU},k}^m \leq w_{\textrm{SU,max}}^m,~ \forall m\in \mathcal{M}\\
\label{SU}\textrm{and,~}w_{\textrm{SU},k}^m=u_k^m(u_k^m-u_{k-1}^m),~\forall k\in \mathcal{K},~\forall m\in \mathcal{M}
\end{eqnarray}
where $w_{\textrm{SU},k}^m$ is a binary variable which is set to 1 if pump $m$ is started up during hour $k$ and $w_{\textrm{SU,max}}^m$ is the maximum number of times a pump $m$ can be started up in a 24-hour period. Equation \eqref{SU} is linearized as follows:
\begin{eqnarray}
\label{pump3}
w_{\textrm{SU},k}^m\leq (u_k^m+1-u_{k-1}^m)/2\\
\label{pump4}
w_{\textrm{SU},k}^m\geq (u_k^m-u_{k-1}^m)/2
\end{eqnarray}

This work considers a total of 7 pump loads - 3 main pumps and 4 auxiliary pumps. In this paper, $Q^m$ = 72,000 $\textrm{m}^3$/h and $w_{\textrm{SU,max}}^m=1$ for all the main pumps; $Q^m$ = 3,600 $\textrm{m}^3$/h and $w_{\textrm{SU,max}}^m=10$ for all the auxiliary pumps and $V_{\textrm{d}}$ = 600,000 $\textrm{m}^3$. The electrical power consumed by each pump during hour $k$ is estimated by the following equation:
\begin{equation}
P^m_k = \beta^mQ^mu_k^m,~\forall k\in \mathcal{K},~\forall m\in \mathcal{M}
\end{equation}
where $P_k^m$ represents the electrical power consumed by pump $m$ during hour $k$ and $\beta^m$ represents the pumping efficiency of pump $m$. In this work, $\beta^m$ = 0.06kWh/$\textrm{m}^3$ for all the main pumps and $\beta^m$ = 0.09kWh/$\textrm{m}^3$ for all the auxiliary pumps.

\subsection{Interruptible Electrical Loads}
Some electrical loads in the system are of a relatively lower priority and can be curtailed if they are adequately compensated. These loads are called interruptible loads (ILs). 
The quantum of IL $h$ curtailed during hour $k$ is constrained as follows:
\begin{eqnarray}
\label{il1}
0 \leq P_{\textrm{EIL},k}^i \leq 2.5\textrm{MWh},~\forall k\in \mathcal{K},~\forall i\in \mathcal{I}\\
\label{il2}
\sum_{i \in \mathcal{I}}P_{\textrm{EIL},k}^i\leq 0.05D_{\textrm{e},k},~\forall k\in \mathcal{K}\\
\label{il3}
\sum_{k\in \mathcal{K}}P_{\textrm{EIL},k}^i \leq 10\textrm{MWh},~\forall i\in \mathcal{I}
\end{eqnarray} 
where $P_{\textrm{EIL},k}^i$ represents the quantum of IL $i$ curtailed during hour $k$; $\mathcal{I}$ represents the set of all ILs in the system and $D_{\textrm{e},k}$ represents the total electrical load demand in the system excluding the flexible pump loads during hour $k$. The total cost incurred by the system operator due to the curtailment of ILs is calculated as follows:
\begin{equation}
C_{\textrm{EIL}} = \sum_{\substack{k\in \mathcal{K}\\i\in \mathcal{I}}} 1.5C_{\textrm{pe},k}P_{\textrm{IL},k}^i
\end{equation}
where $C_{\textrm{pe},k}$ is the price (\$/MWh) at which electrical power is purchased from the utility grid during hour $k$. Three ILs (IL1, IL2 and IL3) characterised by \eqref{il1} - \eqref{il3} are considered in this paper.

\subsection{Flexible Thermal Loads}
A certain percentage of the thermal load demand during each hour is considered to be reschedulable. The usage of the flexible thermal load in the system is constrained as follows:
\begin{eqnarray}
\label{ftl1}
P_{\textrm{Dh},k}=(1-DR^{\textrm{h}}_k)P_{\textrm{Dh},k}^{\textrm{0}}+P_{\textrm{Shift},k}^{\textrm{h}},\forall k \in \mathcal{K}\\
\label{ftl2}
0\leq DR^{\textrm{h}}_k \leq 0.1\\
\label{ftl3} 	
0\leq P_{\textrm{Shift},k}^{\textrm{h}}\\
\label{ftl4}
\sum_{k\in \mathcal{K}}P_{\textrm{Dh},k} = \sum_{k\in \mathcal{K}}P_{\textrm{Dh},k}^{\textrm{0}}
\end{eqnarray}
where $DR^{\textrm{h}}$ represents the percentage of the nominal thermal load $P_{\textrm{Dh},k}^{\textrm{0}}$ which is rescheduled during hour $k$ and $P_{\textrm{Shift},k}^{\textrm{h}}$ represents the thermal load which has been shifted to the current hour $k$ from another hour.

\subsection{Mixed Logical Dynamical Modelling Approach}
Several subclasses of hybrid dynamical systems exist in the literature (see \cite{Hee01} and the references therein). The equivalences between these subclasses were explored in \cite{Hee01}. The MLD formalism is one such subclass which has been used in this paper for modelling the CCPPs, BESS, electrical power interchange with the utility grid, TESSs and boilers. The following equations are used to describe a system in the MLD framework \cite{Bempo10}:
\begin{eqnarray}
\label{e6} 
x(k+1) =  Ax(k) + B_{\textrm{u}}u(k) + B_{\textrm{aux}}w(k) + B_{\textrm{aff}}\\
\label{e8}
E_{\textrm{x}}x(k) + E_{\textrm{u}}u(k) + E_{\textrm{aux}}w(k)\leq E_{\textrm{aff}}
\end{eqnarray}
where $x = [x_{\textrm{c}}\hspace{1mm}x_{\textrm{b}}]^{\textrm{T}}$, $x_{\textrm{c}}\in\mathbb{R}^{n_{\textrm{x}}^{\textrm{c}}}$, $x_{\textrm{b}}\in\{0,1\}^{n_{\textrm{x}}^{\textrm{b}}}$ represents the continuous and binary system states; $u = [u_{\textrm{c}}\hspace{1mm}u_{\textrm{b}}]^{\textrm{T}}$, $u_{\textrm{c}}\in\mathbb{R}^{n_{\textrm{u}}^{\textrm{c}}}$, $u_{\textrm{b}}\in\{0,1\}^{n_{\textrm{u}}^{\textrm{b}}}$ represents the continuous and binary system inputs and $w = [w_{\textrm{c}}\hspace{1mm}w_{\textrm{b}}]^{\textrm{T}}$, $w_{\textrm{c}}\in\mathbb{R}^{n_{\textrm{w}}^{\textrm{c}}}$, $w_{\textrm{b}}\in\{0,1\}^{n_{\textrm{w}}^{\textrm{b}}}$ represents the continuous and binary auxiliary variables. Auxiliary variables are used in the MLD framework to convert propositional logic to linear inequalities of the form \eqref{e8} \cite{Bempo10}. $A$, $B_{\textrm{u}}$, $B_{\textrm{aux}}$, $B_{\textrm{aff}}$, $E_{\textrm{x}}$, $E_{\textrm{u}}$, $E_{\textrm{aux}}$ and $E_{\textrm{aff}}$ are constant matrices of suitable dimensions which are used to define the interactions between the system states, system inputs and auxiliary variables. A detailed description of the MLD framework can be found in \cite{Bempo10}. 

Hybrid system description language (HYSDEL) \cite{Torr04} was used in this paper to develop all the system component models in the MLD framework. The HYSDEL compiler generates all the constant matrices of the MLD model described in (\ref{e6})-(\ref{e8}) from a high-level description of the system behaviour. 
Individual HYSDEL slave files were used to model each GT, ST, boiler, BESS and TESS. The system model was generated by combining individual slave files using the MODULE section of HYSDEL, thereby forming a master file. 
The interactions between the system components were defined in the master file. 
The authors' earlier works \cite{Asho01,KRISHNAN20179329} provide further details on the modelling of CCPPs in the MLD framework. Furthermore, \cite{Pari07} details the modelling of BESSs in the MLD framework.

\section{Optimal Scheduling Problem Formulation} \label{OptSch}
This section describes the formulation of the optimal multi-energy scheduling problem solved in this paper.
Optimal schedules are generated for all the system components described in Section II. The optimal, day-ahead multi-energy scheduling problem is formulated to satisfy all the electrical and thermal loads in the system while respecting various technical and operational constraints described in Section II and later in this section.
Point forecasts for the thermal load demand, electrical load demand, RES generation and utility grid prices for buying/selling electricity are provided as inputs to the different optimal scheduling problems solved in this paper. The following paragraphs describe the hitherto unexplained terms of the objective function.



%
%

$C_\textrm{Fuel}$ represents the cost incurred due to the consumption of natural gas by the GTs in the system. The fuel cost is formulated as a quadratic function of the electrical power produced by the GT.
\begin{equation}
\label{e13}
C_{\textrm{Fuel}} = \sum_{\substack{k\in \mathcal{K}\\f\in GT}}x_{\textrm{disp},k}^f\left(c_{\textrm{2}}^f\left(P_{\textrm{e},k}^f\right)^2  + c_{\textrm{1}}^f P_{\textrm{e},k}^f + c_{\textrm{0}}^f\right)
\end{equation}
where $c_{\textrm{2}}^f$, $c_{\textrm{1}}^f$ and $c_{\textrm{0}}^f$ are the fuel cost curve coefficients of GT $f$ in \$/$\textrm{MW}^{\textrm{2}}$, \$/MW and \$ respectively. 

$C_{\textrm{SU}}$ evaluates the cost incurred during the start-up of all the GTs, STs and boilers in the system. Variable costs are used for the hot, warm and cold start-up methods as shown below. 
\begin{align}
\small
C_{\textrm{SU}} =  \sum_{\substack{k\in \mathcal{K}\\f\in\{GT,ST,BR\}}}  \left( C_{\textrm{cold}}^f\left(w_{\textrm{synch},k}^{\textrm{cold},f}+w_{\textrm{soak},k}^{\textrm{cold},f}\right) \right.+ \hspace{0.7cm} \nonumber \\ 
                   \left. C_{\textrm{warm}}^f\left(w_{\textrm{synch},k}^{\textrm{warm},f}+w_{\textrm{soak},k}^{\textrm{warm},f}\right) \right. \left. +~C_{\textrm{hot}}^fw_{\textrm{soak},k}^{\textrm{hot},f} \right)
\normalsize
\end{align}
where $C_{\textrm{cold}}^f$, $C_{\textrm{warm}}^f$ and $C_{\textrm{hot}}^f$ are the cost coefficients of unit $f$ for cold, warm and hot start-up methods respectively in \$. 

$C_{\textrm{SD}}$ evaluates the cost incurred during the shutdown process of all the GTs, STs and boilers in the system. $C_{\textrm{SD}}$ is calculated as follows:
\begin{equation}
\label{shutdown}
C_{\textrm{SD}} = \sum_{\substack{k\in \mathcal{K}\\f\in\{GT,ST,BR\}}}C_{\textrm{sd}}^fw_{\textrm{desyn},k}^f
\end{equation}
where $C_{\textrm{sd}}^f$ is the shutdown cost coefficient of unit $f$ in \$. 

$C_{\textrm{UCC}}$ is the uncontracted capacity cost. The uncontracted capacity is calculated as follows:
\begin{equation}
\label{UC}
P_{\textrm{UC}} = \max\{0,\underset{1\leq k\leq 24}{\max}\{P_{\textrm{eb},k} - P_{\textrm{CC}}\}\}
\end{equation}
where $P_{\textrm{UC}}$ is the uncontracted capacity in MW and $P_{\textrm{CC}}$ is the contracted capacity in MW. 
Equation \eqref{UC} is linearized as follows:
\begin{eqnarray}
\label{ucc1}
P_{\textrm{UC}}\geq P_{\textrm{eb},k} - P_{\textrm{CC}},~\forall k \in \mathcal{K} \\
\label{ucc2}
P_{\textrm{UC}} \geq 0\\
\textrm{and}~C_{\textrm{UCC}} = U_{\textrm{CC}}P_{\textrm{UC}}
\end{eqnarray}
where $U_{\textrm{CC}}$ = \$12,860/MW/month and $P_{\textrm{CC}}$ = 25 MW.

$C_{\textrm{Boiler}}$ evaluates the boiler fuel cost. It is assumed that all the boilers modelled in this paper use natural gas as fuel to produce thermal energy. The natural gas price is considered to be \$3.81/mcf in this paper.
\begin{equation}
C_{\textrm{Boiler}} = \sum_{\substack{k\in \mathcal{K}\\f\in BR}}3.81w_{\textrm{br},k}^f
\end{equation}

$C_{\textrm{Grid}}$ accounts for the cost incurred due to the purchase of electrical and thermal power from external sources. $C_{\textrm{Grid}}$ also includes the revenue earned from the sale of electrical power to the main utility grid. $C_{\textrm{Grid}}$ is calculated as follows:
\begin{equation}
C_{\textrm{Grid}} = \sum_{k\in \mathcal{K}} \left(C_{\textrm{p},k}P_{\textrm{eb},k} - C_{\textrm{s},k}P_{\textrm{es},k} + C_{\textrm{heat}}P_{\textrm{hb},k}\right)
\end{equation}
where $C_{\textrm{p},k}$ is the price at which electrical power is purchased from the main utility grid during hour $k$; $P_{\textrm{es},k}$ is the electrical power sold to the main utility grid during hour $k$ in MW; $P_{\textrm{hb},k}$ is the thermal power purchased from external sources during hour $k$ in MW and $C_{\textrm{s},k}$ is the price at which electrical power is sold to the main utility grid during hour $k$. Finally, $C_{\textrm{heat}}$ = \$100/MW is the price at which thermal power is purchased from external sources. 

The overall optimal scheduling problem for the multi-energy system described in this paper is summarized as follows:
\begin{align}
\underset{u, x, w} {\min}~J = C_{\textrm{Fuel}} + C_{\textrm{BESS}} + C_{\textrm{SU}} +  C_{\textrm{SD}} + C_{\textrm{UCC}} + C_{\textrm{Boiler}}  \hspace{0.2cm}  \nonumber \\ 
+~C_{\textrm{Grid}} + C_{\textrm{EIL}} \hspace{4.3cm} \label{cost_fun} \\
 {\textrm{subject to}} \hspace{0.4cm}\textrm{Equations}~\eqref{pump1},\eqref{pump2},\eqref{pump3},\eqref{pump4}, \eqref{il1}-\eqref{il3}, \hspace{0.5cm} \nonumber \\ 
    \eqref{e6},\eqref{e8},\eqref{ucc1},\eqref{ucc2},\eqref{ftl1}-\eqref{ftl4}  \hspace{2cm} \nonumber \\
\small
P_{\textrm{De},k} + \sum_{m\in \mathcal{M}}P_k^m - \sum_{h \in \mathcal{H}}P_{\textrm{EIL},k}^h \leq  \sum_{f\in\{GT,ST\}} \left(P_{\textrm{e},k}^f + P_{\textrm{soak},k}^f\right) \nonumber \\ 
\normalsize
+ P_{\textrm{eb},k} - P_{\textrm{es},k} + P_{\textrm{bd},k} - P_{\textrm{bc},k} + P_{\textrm{RES},k} \label{heat_bal} \\
P_{\textrm{Dh},k}+\sum_{f\in ST}h_k^f \leq \sum_{f\in\{GT,BR\}}  \left(P_{\textrm{h},k}^f\right) + P_{\textrm{hb},k} - Q_{\textrm{in},k}^1 \hspace{0.8cm} \nonumber \\ 
-Q_{\textrm{in},k}^2+Q_{\textrm{out},k}^1+Q_{\textrm{out},k}^2 \hspace{0.6cm}  \label{elec_bal} \\
u_{\textrm{min}}\leq u(k) \leq u_{\textrm{max}}\\x_{\textrm{min}}\leq x(k) \leq x_{\textrm{max}}\\ w_{\textrm{min}}\leq w(k) \leq w_{\textrm{max}}\hspace{0.2cm}  \hspace{0.5cm} \label{bounds}\\
0\leq P_{\textrm{eb},k} \leq 50,~0\leq P_{\textrm{eb},k}\leq 50,~0\leq P_{\textrm{hb},k}  \leq 80 \hspace{1cm}
\end{align}
where $P_{\textrm{RES},k}$ represents the combined electrical power produced by the two wind power plants in the system during hour $k$. Furthermore, \eqref{heat_bal} and \eqref{elec_bal} represent the electrical and thermal power balance constraints respectively.
The overall optimization problem turns out to be an MIQP problem which is formulated in MATLAB using YALMIP \cite{Lof14} and solved using CPLEX. For the sake of brevity, the technical parameters of all the GTs, STs and boilers modelled in this work are provided at \url{http://dx.doi.org/10.13140/RG.2.2.28684.21122}.

\section{Case Studies}
To demonstrate the efficacy of the optimal multi-energy scheduling problem formulated in Section III, the following scenarios are simulated:
\begin{enumerate}
	\item Load scheduling is not performed. The liquid is pumped out in the fastest possible time using only the main pumps. The auxiliary pumps, flexible thermal load and ILs are not included in the optimal scheduling problem formulation for this scenario while the schedules of the main pumps are fixed. The electrical and thermal load demand is entirely made up of critical loads.
	\item Load scheduling is performed to demonstrate the flexibility offered by the PSO. All the main pumps and the auxiliary pumps participate in the PSO. The flexible thermal load and ILs are not included in the optimal scheduling problem formulation for this scenario.
	\item In addition to the PSO, this scenario considers the presence of ILs which relaxes the optimal scheduling problem and provides further flexibility to the system operator. The flexible thermal load is not included in the optimal scheduling problem formulation for this scenario.
	\item In addition to the PSO and ILs, the flexible thermal load is included in the optimal scheduling problem formulation for this scenario. This scenario truly represents the optimal scheduling problem formulation presented in Section III. As demonstrated later in this section, this scenario offers the maximum flexibility to the system operator, thereby resulting in the lowest energy cost among all the simulated scenarios.
\end{enumerate}

\subsection{System Initialization}
Initially, it is assumed that GT1, GT2, ST1, ST2, ST3, Boiler 1 and Boiler 2 are already in the dispatch phase. Furthermore, $SOC_{1}$ = 0.6 and $H_1^1$ = $H_1^2$ = 171.643MW. All the main and auxiliary pumps are assumed to be in the OFF position prior to the start of the optimization period. The initial system states have been carefully chosen to ensure a feasible operating point for the system prior to the start of the optimization period. It is also pertinent to mention here that the initial states of the system have a significant bearing on the final system trajectory and the scheduling results obtained. However, the system initialization does not significantly alter the general trends observed in the results presented later in this section.

\subsection{Results and Discussions}
\begin{figure}[h!]
	\begin{center}
		\includegraphics[width=3.5in]{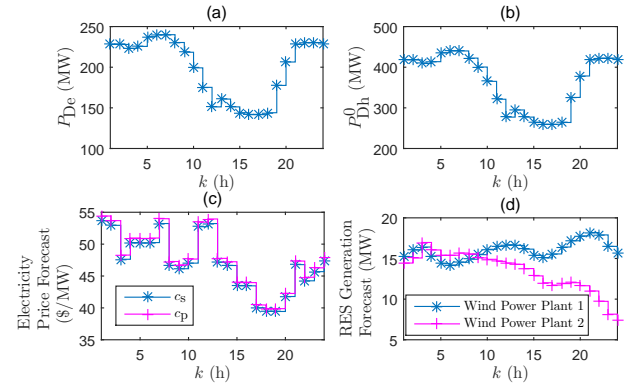}
		\caption{Point forecasts for: (a) $P_{\textrm{De}}$ (b) $P_{\textrm{Dh}}^{\textrm{0}}$ (c) $c_{\textrm{s}}$ and $c_{\textrm{p}}$ and (d) RES generation}
		\label{opt_inputs}  
	\end{center}
\end{figure}

\begin{figure}[h!]
	\begin{center}
		\includegraphics[width=3.5in]{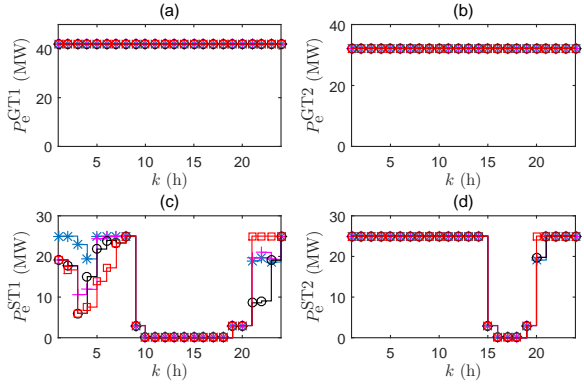}
		\caption{Electrical power dispatch values under Scenarios 1-4 of: (a) GT1 (b) GT2 (c) ST1 and (d) ST2. The legend for (a), (b), (c) and (d) is as follows: Scenario 1 - blue *, Scenario 2 - magenta +, Scenario 3 - black circle and Scenario 4 - red square.}
		\label{subplot1}  
	\end{center}
\end{figure}

\begin{figure}[h!]
	\begin{center}
		\includegraphics[width=3.5in]{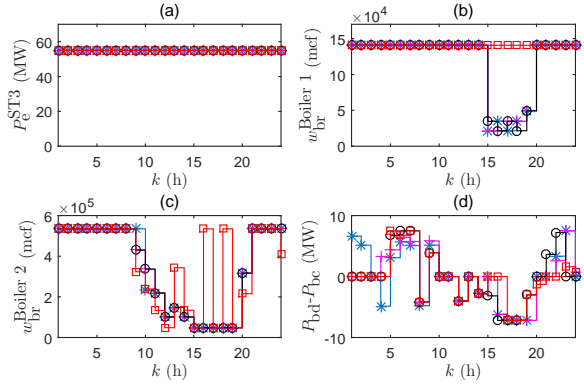}
		\caption{Profiles (under Scenarios 1-4) of: (a) Electrical power dispatch of ST3 (b) Fuel consumption of Boiler 1 (c) Fuel consumption of Boiler 2 and (d) BESS usage represented by $P_{\textrm{bd}}-P_{\textrm{bc}}$. The legend for (a), (b), (c) and (d) is as follows: Scenario 1 - blue *, Scenario 2 - magenta +, Scenario 3 - black circle and Scenario 4 - red square.}
		\label{subplot2}  
	\end{center}
\end{figure}

\begin{figure}[h!]
	\begin{center}
		\includegraphics[width=3.5in]{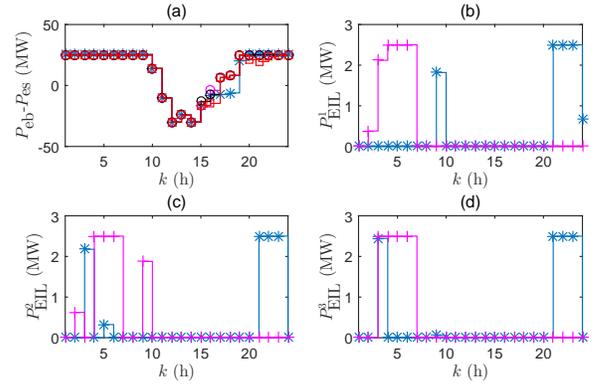}
		\caption{Profiles (under Scenarios 1-4) of: (a) Electricity exchanged with the main grid represented by $P_{\textrm{eb}}-P_{\textrm{es}}$ (b) Usage of IL1 (c) Usage of IL2 and (d) Usage of IL3. The legend for (a) is as follows: Scenario 1 - blue *, Scenario 2 - magenta +, Scenario 3 - black circle and Scenario 4 - red square. The legend for (b), (c) and (d) is as follows: Scenario 3 - blue * and Scenario 4 - magenta +.}
		\label{subplot3}  
	\end{center}
\end{figure}

\begin{figure}[h!]
	\begin{center}
		\includegraphics[width=3.5in]{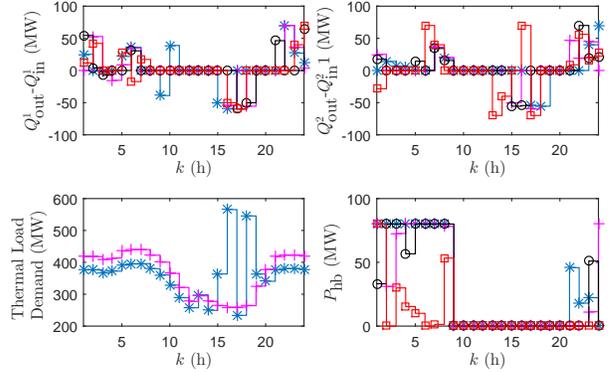}
		\caption{Profiles (under Scenarios 1-4) of: (a) Usage of TESS 1 (b) Usage of TESS 2 and (c) $P_{\textrm{Dh}}$ and $P_{\textrm{Dh}}^{\textrm{0}}$. The legend for (a), (b) and (d) is as follows: Scenario 1 - blue *, Scenario 2 - magenta +, Scenario 3 - black circle and Scenario 4 - red square. The legend for (c) is as follows: $P_{\textrm{Dh}}$ - blue * and $P_{\textrm{Dh}}^{\textrm{0}}$ - magenta +.}
		\label{subplot4}  
	\end{center}
\end{figure}

The inputs to the optimal scheduling problem are shown in Fig. \ref{opt_inputs}(a) - Fig. \ref{opt_inputs}(d). The point forecasts for the electrical and thermal load demands are shown in Figs. \ref{opt_inputs}(a) and \ref{opt_inputs}(b) respectively. The point forecasts for the electricity price (obtained from \cite{emc}) and RES generation are shown in Figs. \ref{opt_inputs}(c) and \ref{opt_inputs}(d) respectively. The results of the optimal scheduling problem solved under all four scenarios are presented in Fig. \ref{subplot1} - Fig. \ref{subplot4} and Tables \ref{t1} and \ref{t2}.

Figs. \ref{subplot1}(a), \ref{subplot1}(b) and \ref{subplot2}(a) indicate that GT1, GT2 and ST3 service the electrical base load demand under all four scenarios. As such, they operate at full capacity throughout the day under all four scenarios. From Figs. \ref{subplot1}(c) and \ref{subplot1}(d), the effect of including the startup/shutdown power trajectories can be clearly observed. From Fig. \ref{subplot1}(c), it is observed that ST1 is unused between hours 10-18 under all four scenarios due to the low electrical load demand during those hours. The pump schedules in Table \ref{t1} also show that the pumps are operated during hours 16-20 in Scenarios 2-4 to avoid uncontracted capacity costs. Fig. \ref{subplot2}(d) indicates that the usage of the BESS follows a similar trend under Scenarios 1-4. By and large, it is observed that the BESS charging takes place during the hours when the electrical load demand is low.

Under Scenario 1, the main pumps are operated during the first 3 hours of the optimization period. From Fig. \ref{subplot1}(c), it is observed that the utilization of ST1 is higher during the first 4 hours under Scenario 1 when compared with the other scenarios. This is to cater to the additional electricity demand caused by the operation of the main pumps during these hours. From Fig. \ref{subplot4}(d), it is observed that the dependence on imported thermal energy is the highest under Scenario 1, especially during the first 8 hours. This is due to the high utilization of the STs coinciding with the high thermal load demand during these hours. As observed in Fig. \ref{subplot3}(a), imported electricity from the main utility grid is used to mitigate any shortfall in the electricity generated within the multi-energy system during the first few hours of the optimization period. This leads to the consumption of uncontracted capacity which entails a huge cost. As observed in Fig. \ref{subplot2}(d), the BESS utilization (in discharging mode) during the first 2 hours is also quite high under Scenario 1. This is to cope with the additional electricity demand during these hours.


Compared with Scenario 1, the PSO performed under Scenario 2 eliminates the uncontracted capacity cost, thereby leading to a reduction in the total energy cost of the system as shown in Table \ref{t2}. As shown in Table \ref{t1}, this is achieved by shifting the usage of the pumps to the off-peak hours (hours 16-19) from the peak hours. Consequently, as observed from Figs. \ref{subplot2}(d) and \ref{subplot1}(b), there is a decrease in the usage of the BESS and ST1 respectively. The reduced usage of ST1 leads to a slight decrease in the requirement of imported thermal energy during the first 8 hours as seen in Fig. \ref{subplot4}(d). There is also a significant quantity of thermal energy imported during hours 21-23 under Scenario 1 and hours 23-24 under Scenario 2. This is to cater to the high thermal load demand experienced during these hours. From Figs. \ref{subplot4}(a) and \ref{subplot4}(b), it is observed that thermal energy is also drawn from the TESSs during these hours under Scenarios 1 and 2. From Figs. \ref{subplot2}(b) and \ref{subplot2}(c), it is seen that both Boiler 1 and Boiler 2 are also operated at full capacity during these hours under Scenarios 1 and 2. The BESS is also used in the discharging mode during hours 22-23 as seen in Fig. \ref{subplot2}(d). From  Fig. \ref{subplot1}(c), it is observed that the utilization of ST1 is lower under Scenario 3 than under Scenario 1 during hours 1-5 and lower than under Scenario 2 during hours 3 and 5. This is mainly due to the utilization of the ILs as observed from Figs. \ref{subplot3}(b) - \ref{subplot3}(d). A similar phenomenon is also observed during hours 21-23 under Scenario 3. During hour 24, only IL1 is utilized under Scenario 3. This leads to an increased utilization of ST1 during hour 24. Furthermore, as seen from Fig. \ref{subplot2}(d), the BESS also discharges during hours 21 and 22 under Scenario 3 to cope with the higher electrical load demand. Under Scenario 3, from Fig. \ref{subplot4}(d), it is observed that thermal energy is imported during hour 23. Furthermore, from Figs. \ref{subplot4}(b) and \ref{subplot4}(c), it is observed that the TESSs supply thermal energy during hours 21-24 under Scenario 3 to cope with the higher thermal load demand.


Under Scenario 4, the purchase of expensive thermal energy from external producers is the least among the four scenarios as observed in Fig. \ref{subplot4}(d). This can be largely attributed to the introduction of the flexible thermal load in the problem formulation for Scenario 4 which causes some of the thermal load demand during the peak load hours to be shifted to the off-peak hours as shown in Fig. \ref{subplot4}(c). For instance, it is observed that the profile of $P_{\textrm{Dh}}$ has distinct spikes during hours 16 and 18. This can be attributed to the shifting of the thermal load to these hours from the peak loading hours. Additionally, unlike the other scenarios, it is observed in Fig. \ref{subplot2}(c) that the usage of Boiler 2 also rises during hours 16 and 18 under Scenario 4 to cater to the additional thermal load demand. Furthermore, from Fig. \ref{subplot2}(b), it is observed that Boiler 1 is also operated at full capacity during the entire optimization period under Scenario 4. From Figs. \ref{subplot3}(b) - \ref{subplot3}(d), it is observed that the ILs are also mainly utilized between hours 2-6 and during hour 9 under Scenario 4 to relax the optimal scheduling problem and to compensate for any shortfall in the electricity production without resorting to uncontracted capacity consumption. The utilization of ST1 during hours 4-6 under Scenario 4 is the lowest among all the four scenarios due to the usage of the ILs during these hours. The combined effect of the ILs and the flexible thermal load causes the electrical and thermal load demands during the peak load (electrical and thermal) hours to reduce, thereby obviating the need to import uncontracted capacity and large quantities of thermal energy. Consequently, Scenario 4 has the lowest energy cost among all the simulated scenarios as seen in Table \ref{t2}. Compared with Scenario 1 (worst-case scenario), the energy cost under Scenario 4 is 18.6\% lower. As the flexibility available to the system operator is progressively increased under Scenarios 2-4, the cost progressively declines. The greater flexibility allows the system operator to better manage the load demand using locally available generation while sparingly resorting to energy imports as and when necessary.

\renewcommand{\arraystretch}{1.5}
\begin{table*}[ht!]
\begin{center}
\caption{Pump schedules under Scenarios 1-4.} \label{t1}
\begin{tabular}{|c|c|c|c|c|}
\hline
Pump No.            &	Scenario 1                     &          Scenario 2        	&          Scenario 3             &      Scenario 4            \\
\hline \hline
Main pump 1	    &   111000000000000000000000       &    000000000000000111100000 	&     000000000000000011100000    &  000000000000000001110000  \\
                     
Main pump 2         &   111000000000000000000000       &    000000000000000011000000    &     000000000000000011100000    &   000000000000000001100000  \\

Main pump 3         &   111000000000000000000000       &     000000000000000011000000   &     000000000000000011000000    &   000000000000000001110000   \\                     

Auxiliary pump 1    &   000000000000000000000000       &     000000000000000001000000   &     000000000000000001000000    &   000000000000000000100000   \\                     

Auxiliary pump 2    &   000000000000000000000000       &     000000000000000011000000   &     000000000000000011000000    &   000000000000000000100000   \\                     

Auxiliary pump 3    &   000000000000000000000000       &     000000000000000011000000   &     000000000000000011000000    &   000000000000000001100000   \\                     

Auxiliary pump 4    &   000000000000000000000000       &     000000000000000011000000   &     000000000000000011000000    &   000000000000000001100000   \\                     
\hline
\end{tabular}
\end{center}
\end{table*}
\renewcommand{\arraystretch}{1.5}

\setlength{\tabcolsep}{5pt}
\begin{table}[]
	\begin{center}
		\caption{Cost comparison under Scenarios 1-4}\label{t2}
		\begin{tabular}{|c|c|c|c|}
		\hline
		\textbf{Scenario} & \textbf{\begin{tabular}[c]{@{}c@{}}Uncontracted\\ Capacity\\ Cost (\$)\end{tabular}} & \textbf{Total Cost (\$)} & \textbf{\begin{tabular}[c]{@{}c@{}}Percentage\\  Reduction\\  with respect to\\  Scenario 1\end{tabular}} \\ \hline
		1                 & 8,558.18                                                                              & 298,822.8                & -                                                                                                         \\ \hline
		2                 & 0                                                                                    & 285,881.83               & 4.33                                                                                                      \\ \hline
		3                 & 0                                                                                    & 282,769.35               & 5.37                                                                                                      \\ \hline
		4                 & 0                                                                                    & 243,183.54               & 18.62                                                                                                     \\ \hline
	\end{tabular}
\end{center}
\end{table}

\subsection{Limitations and Scope for Future Work}
The results presented in this section clearly demonstrate the efficacy and cost reduction potential of the optimal scheduling model presented in this paper. However, the model presented in this paper does have its limitations, thereby opening several areas for future research. Firstly, the model presented in this work does not consider any electrical and thermal network constraints which could potentially affect the feasibility of the schedule generated for the system. Including the electrical and thermal network constraints in the optimal scheduling model is an area of ongoing research. The other major direction for future research is the consideration of uncertainties in the RES generation, load demand and electricity price forecasts. This would require the adoption of advanced optimization procedures such as stochastic and robust optimization techniques. Finally, the model presented in this work considers linear relationships to describe the heat generated by the GTs and the heat consumed by the STs. The adoption of more accurate models to describe these relationships in the optimal scheduling model is also an interesting area for future research.

\section{Conclusion}
This paper presented an optimal, day-ahead scheduling model for an exemplar multi-energy system comprising CCPPs, boilers, RESs, BESS, TESSs, flexible thermal load, flexible pump loads and ILs. The multi-energy system model presented in this paper included a detailed treatment of the startup and shutdown power trajectories inherent to the GTs, STs and boilers. A major part of the system model was constructed using the MLD modelling framework. Furthermore, a multi-energy load management scheme was included in the optimal scheduling model. The multi-energy load management scheme took advantage of the flexibility offered by PSO, flexible thermal load and ILs to drive down the energy cost of the system. The efficacy and cost reduction potential of the optimal scheduling model was demonstrated using four illustrative simulation scenarios. The best-case scenario delivered an 18.6\% cost reduction when compared with the worst-case scenario. The simulated scenarios were analysed to demonstrate how the optimization model aided in reducing the energy cost of the system. Finally, the limitations of this work and the consequent directions for future research were also presented.

%
\section*{Acknowledgment}
The authors acknowledge funding support from NTU Start-Up Grant. The authors also acknowledge the support for research from the National Research Foundation, Prime Ministers Office, Singapore under its Campus for Research Excellence and Technological Enterprise (CREATE) programme.

\bibliographystyle{ieeetr}
\bibliography{references}

\end{document}